\documentclass[a4paper]{jpconf}
\usepackage{graphicx,epsfig}
\newcommand{\be}{\begin{eqnarray}}
\newcommand{\ee}{\end{eqnarray}}

  \newcommand{\lqcd}{\Lambda_{\mathrm{QCD}}}
\newcommand{\eqcomma}{\phantom{AA},\phantom{AA}}
\newcommand{\order}[1]{ \mathcal{O} \left( #1 \right) }

\begin{document}
\title{Phenomenology of quarkyonic percolation at FAIR}
\author{Giorgio Torrieri, Stefano Lottini}
\address{FIAS, J.~W.~Goethe Universit\"at, Frankfurt Am Main, Germany.}
\begin{abstract}
We will give an introduction to the concept of quarkyonic matter, presenting an overview of what is meant by this term in the literature.  We will then argue that the quarkyonic phase, as defined in the original paper, is a percolation-type phase transition whose phase transition line is strongly curved in $\rho_B-N_c$ space, where $N_c$ is the number of colors and $\rho_B$ the baryon density.   With a  toy model estimate, we show that it might be possible to obtain a percolating but confined phase at $N_c=3,N_f=2$ at densities larger than one baryon per one baryon size.   We conclude by discussing how this phase can be observed at FAIR.
\end{abstract}
A concept recently receiving quite a lot of theoretical attention is that of ``quarkyonic matter''.   Introduced in \cite{quarkyonic}, it can be best defined as matter where temperature is below deconfinement, $T < T_c$, but quark chemical potential is of the order of the confinement scale (inverse  of the baryon size), $\mu_Q \geq \lqcd$, resulting in a density of a baryon per baryonic volume or higher.
This regime is not easily accessible through controlled theoretical expansions:
Effective hadronic models, since they expand in powers of $Q/\lqcd$, where $Q$ is the momentum exchange, can be reasonably expected to diverge.   Lattice calculations are affected by the sign problem, which generally can only be controlled at $\mu_Q \ll T$.

Hence, expectations for matter in this regime are very model-specific.
Popular ideas included a confined yet chirally restored phase \cite{quarkyonic5,quarkyonic2}, chirally inhomogeneus phases \cite{carignano,spiral,quarkyonic6}, a deconfined but chiral-broken plasma of constituent quarks \cite{satzquarkplasma} and a large-$N_c$ generalization of nuclear matter \cite{lippert,adsquark,usquark}.
\begin{figure}[h]
\begin{center}
\epsfig{width=13cm,figure=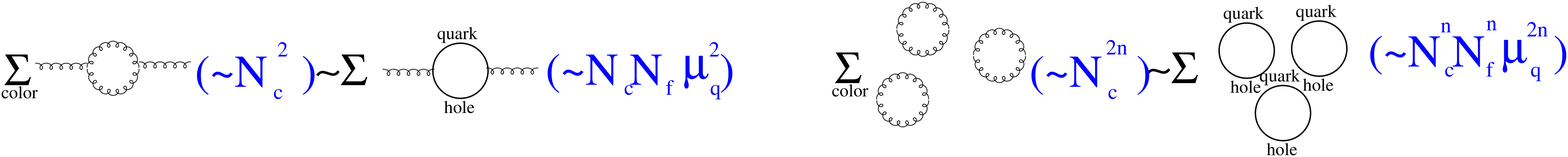}
\caption{\label{decodiag} Diagrams at arbitrary $N_c$ responsible for gluon antiscreening and quark-hole screening.}
\end{center}
\end{figure}
Ref.~\cite{quarkyonic} has generically argued that in this regime confinement coexists with pQCD and asymptotic freedom.
The essence of this argument is rather simple and qualitative: 
Assuming that at finite chemical potential deconfinement appears when the quark-hole screening overpowers the gluon-gluon antiscreening, we can obtain how the chemical potential required for deconfinement varies with $N_c$.
It turns out that at leading order (Fig.~\ref{decodiag}), $\mu_Q^{deconfinement} \simeq \sqrt{N_c/N_f} \lqcd$.  At higher order ($n$ loops) the $N_c$ dependence is parametrically unchanged.
Thus, looking at momentum space, we expect $\mu_Q \simeq \lqcd$ matter to be firmly confined.

In configuration space, on the other hand, we have a baryon with $N_c$ quarks per baryonic size.   Hence, quarks of neighboring baryons are distant by $N_c^{-1/3} \rightarrow 0$.   Thus, at least at large $N_c$, asymptotic freedom should somehow coexist with confinement.   A natural picture of this, found in \cite{quarkyonic}, is an asymptotically free gas of quarks moving in a mean field of baryons.   Alternatively, the bulk of the Fermi surface is asymptotically free but its excitations are confined.  
\begin{figure}[h]
\begin{center}
\epsfig{file=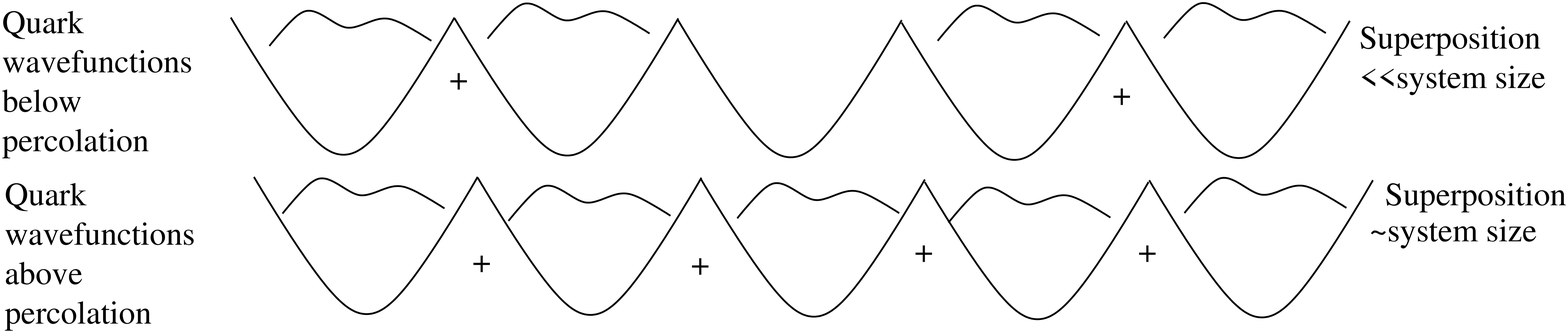,width=12cm}
\caption{\label{percchain} The influence of percolation of quark wavefunctions inside a dense distribution of hadrons (shown here as potential wells).}
\end{center}
\end{figure}

How relevant is this reasoning to our $N_c=3,N_f=2$ world?
  As we have already known for a few decades, the $N_c \rightarrow \infty$ expansion is qualitatively remarkably good for a lot of QCD phenomenology:  That mesons are weakly-interacting quasi-particles, baryons look like semi-classical ``skyrmion'' states of mesonic fields, and planar diagrams dominate are all features of the large $N_c$ expansion.
On the other hand, some features of QCD thermodynamics are not even qualitatively well reproduced:  At large $N_c$, deconfinement is a first order phase transition rather than a cross-over.   At large chemical potential, binding energy of nuclear matter is {\em of the order of the baryon mass} $\sim N_c \lqcd$ (rather than $\sim 10^{-3} N_c \lqcd \ll m_\pi$, as in our world).

Thus, the existence of phase transitions in $N_c$, as well as $T$ and $\mu_Q$, is virtually assured.   Such transitions can not obviously be studied through experimental data, but nevertheless remain pretty interesting theoretically: The large-$N_c$ technique is one that has been extensively used for phenomenology as well as theory.    The identification of the large-$N_c$ limit with a classical dual gauge theory makes this topic more interesting still, since the phase transition in $N_c$ is a quantum-gravitational one in its dual description.    Unfortunately, as confinement shows, determining that a phase transition exists from a fundamental theory can be a very subtle issue.

\begin{figure}[h]
\begin{center}
\epsfig{file=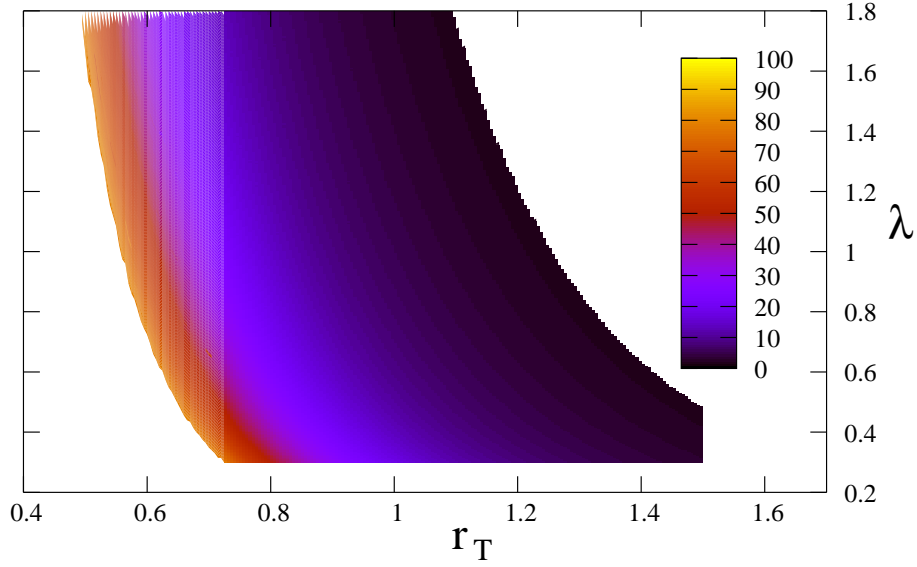,width=12cm}
\caption{\label{perc1} The dependence of the percolation critical number of colors $N_c$ on $\lambda,r_T$ at a density of one baryon per baryon size and hexagonal packing \cite{perc1}.}
\end{center}
\end{figure}
In the rest of this work, we shall use percolation \cite{perc1,perc2} as a ``toy model'' to investigate a transition between normal nuclear matter and the quarkyonic phase which bears a lot of similarity with that described in \cite{quarkyonic}.   Percolation is a simple yet surprisingly deep and universal paradigm:  It requires an infinite network of ``links'', each with a probability, invariant in space, to be ``on'' or ``off''.
The crucial question is the typical length of a connected cluster of ``on'' links as a function of the individual link probability.\footnote{Excluding the infinite component, if any.}
Somewhat subtly, this link's dependence on probability is not monotonic: At the critical ``percolation'' probability (typically not very high, $\order{1}/N_{neighbours}$), this length diverges.    Considering this as a correlation length, it is clear percolation defines a second order phase transition, an insight which is confirmed by the appearance of critical behavior at the percolation threshold.
\begin{figure}[h]
\epsfig{file=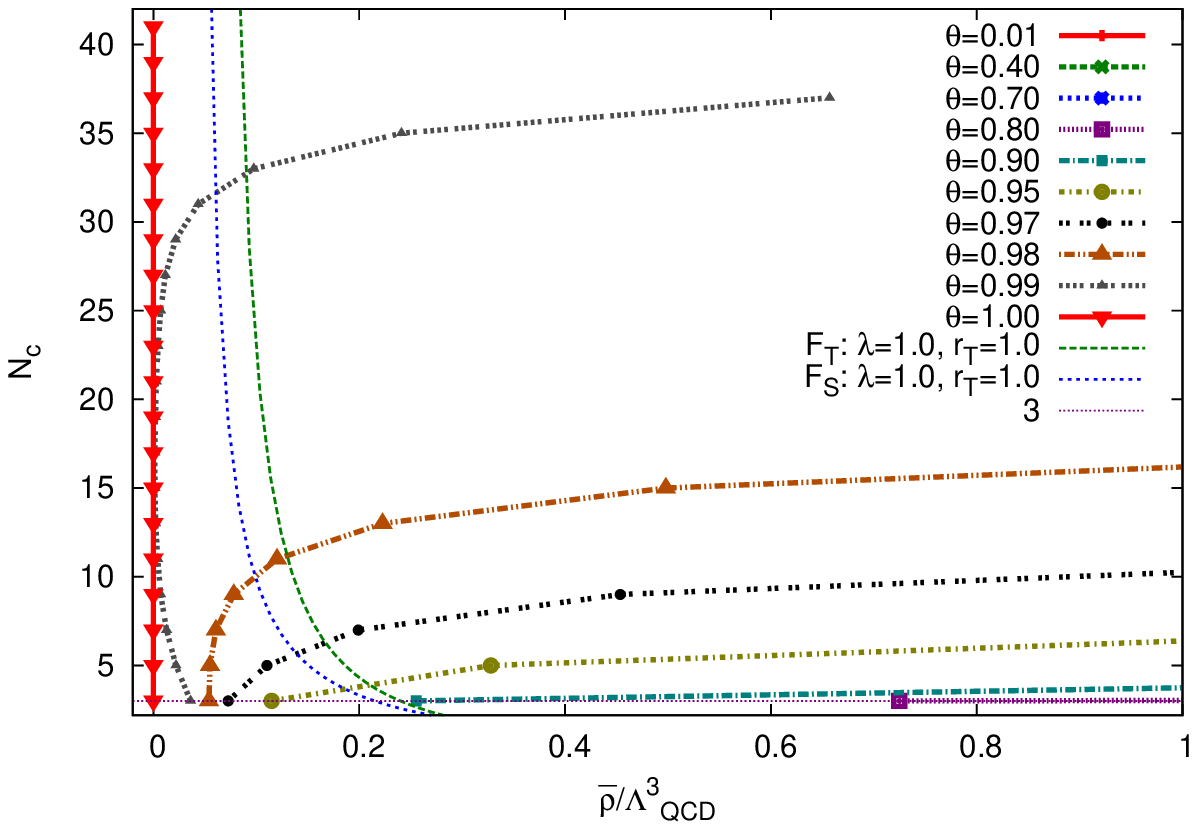,width=7cm}
\epsfig{file=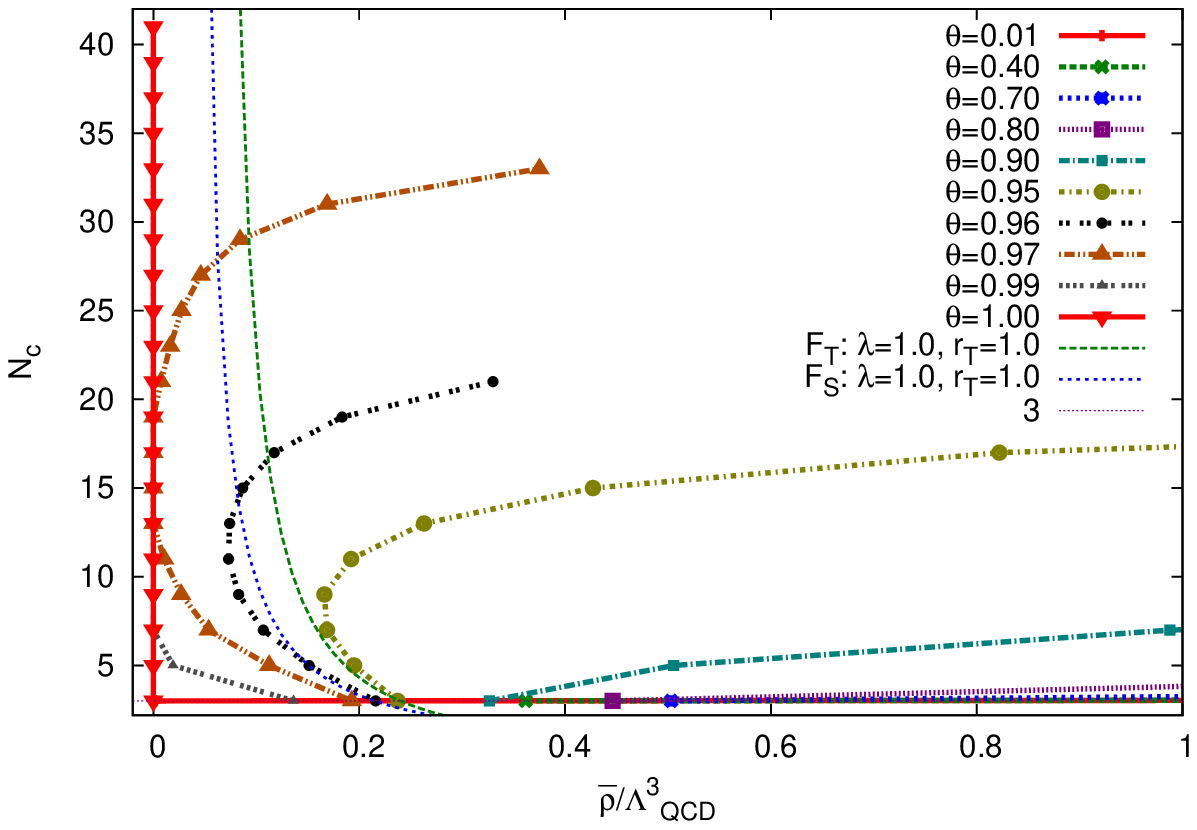,width=7cm}
\caption{\label{perc2}Two-dimensional plots in the $($baryon density,$N_c)$ plane of the percolation curves (lines without data points) vs.~the deconfinement curves at various temperatures ($\theta=T/T_c$), for $N_f=1,3$ (resp.~left and right plots).}
\end{figure}
The abstractness and generality of percolation makes it very appealing as a toy model, since {\em any} local propagator in an extended medium should give rise to percolation.
 Specifically, by analogy with the well-known conductor-insulator transition in QED, percolation signals the regime where one can not consider quark wavefunctions to be effectively localized to a baryon since quantum tunneling will have the potential to make quarks propagate through arbitrary distances at equilibrium (Fig.~\ref{percchain}).

To investigate the curvature in $\rho_B$-$N_c$ space, we posutulate a generic interaction propagator, respecting what we know about Yang-Mills theory:  Due to confinement, it has to fall off at distances higher than $\lqcd^{-1}$ in configuration space, and the interaction probability has to $\sim \lambda/N_c$:
\begin{equation}
f(x) = \frac{\lambda}{N_c} f\left( \frac{x}{r_T} \right) \eqcomma  f\left( \frac{x}{r_T} \geq 1 \right) \rightarrow 0\;\;\;;
\end{equation}
in units of $\lqcd$, it is natural that $r_T \sim \order{1}$.  Since at that scale $\lambda$ becomes strong, it is natural that that it is $\lambda \sim \order{1}$ too.   As long as these requirements are satisfied, the shape of $f(x)$ ($\Theta$-function in position space, Gaussian, etc) is irrelevant.
\begin{figure}[h]
\begin{center}
\epsfig{width=14cm,figure=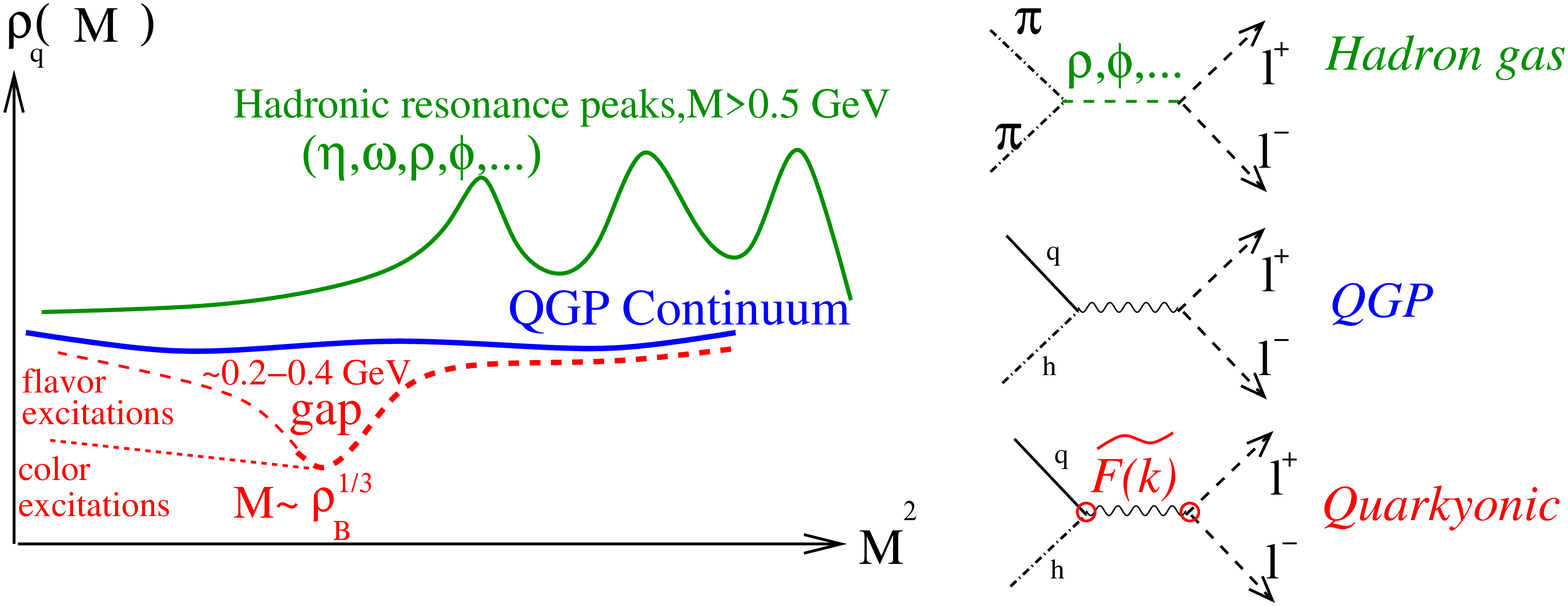}
\caption{\label{figee} Spectral function of dileptons in a hadron gas, in a pQCD, and in quarkyonic matter.
}
\end{center}
\end{figure}
Fig.~\ref{perc1} \cite{perc1} shows the dependence of the critical number of colors $N_c$ for a regular hexagonal packing on $\lambda,r_T$ at fixed baryon density.
It is clear that at the chemical potential of $\mu_Q \simeq \lqcd$, indicated in \cite{quarkyonic}, $N_c=3$ is not enough to trigger a percolation transition, which requires $N_c \simeq \order{10}$.   However, the possibility of quarkyonic matter at $1 \leq \mu_Q/\lqcd \leq \sqrt{N_c/N_f}$ remains open if the effective number of flavors $N_f <3$.   Note that, since the mass of the strange quark $\simeq \lqcd$, it is far from clear whether the above hyerarchy can be maintained, since it is uncertain whether the strange flavor is ``light'' for our purposes.

To further investigate whether the quarkyonic phase can appear, we varied density as well as $N_c$ \cite{perc2}.   Note that, while this is an obvious extension, technically it is somewhat difficult because the definition of ``nearest neighbor'' is somewhat sloppy with varying density (the nearest neighbor at one density can be farther than the next-to-nearest neighbor at the other).
To obtain the critical percolation probability, we integrated out physical baryons into a square lattice, and performed a renormalization-group analysis.  The details of this procedure are available in \cite{perc2}.

 Unlike percolation, baryonic density necessary for deconfinementgoes up with $N_c$ because of the $\mu_Q \sim \sqrt{N_c/N_f} \lqcd$ requirement.
Hence, the ``true high-$N_c$ limit'' is one where deconfinement happens at higher $\mu_Q$ than percolation.   In this limit, there will be both a confined percolating phase and a deconfined phase.   At the low $N_c$ limit, percolation would require a higher $\mu_Q$ than deconfinement.   Since percolation requires baryonic wavefunctions to be present, this means percolation {\em never} occurs.
 Where is our world in this hyerarchy? In Fig.~\ref{perc2} \cite{perc2} we plotted, as a function of $N_c,\mu_Q,T$, the density of an ideal baryon gas at deconfinement.   As can be seen, a sliver of percolating but confined phase diagram remains.    While the ``ideal'' approximation for baryon gas density is rough in this limit, this calculation shows percolation is {\em possible} in our world.   Hence, an experimental search for it might be necessary.
\begin{figure}[h]
\begin{center}
\epsfig{width=14cm,figure=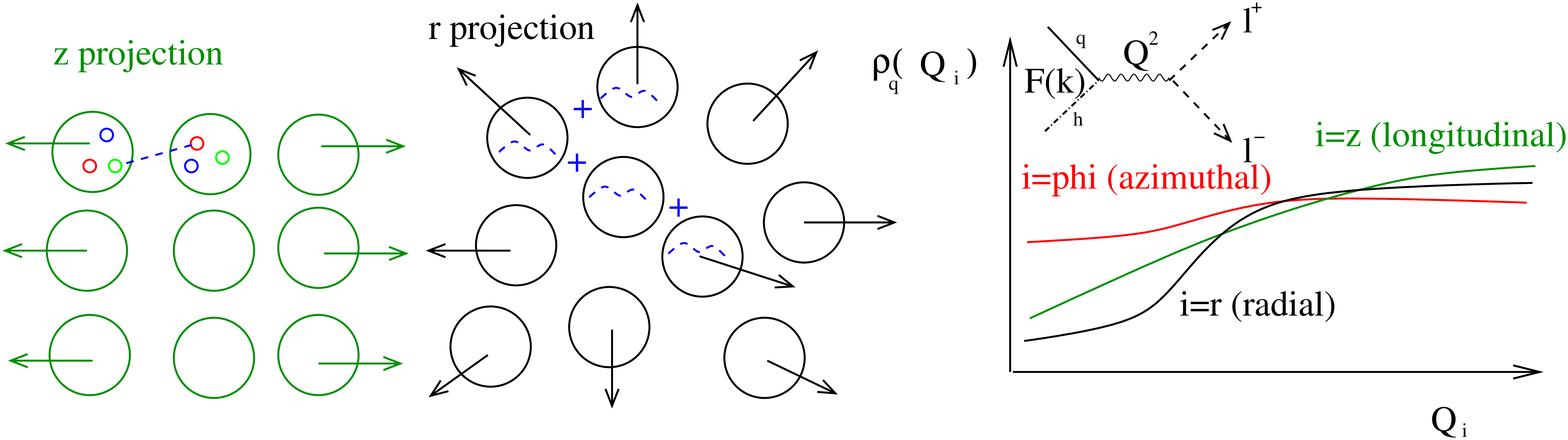}
\caption{\label{figee_azimuth} Influence of the baryonic dynamics on the decomposed quark-hole spectral function in quarkyonic matter.}
\end{center}
\end{figure}

What would be an experimental signature for the production of ``quarkyonic matter''?   This question was not really answered by \cite{quarkyonic};  however, the coexistence, in quarkyonic matter, of pQCD degrees of freedom with baryonic mean fields makes exploration of the spectral function by dilepton detection a promising avenue to investigate.
Consider a reaction such as
 $
q + \mathrm{hole} \rightarrow \gamma^* \rightarrow l^+ l^-
$
whose rate $\Gamma$ will be given by
\begin{equation}
\Gamma(M) = \int \mathcal{M}\left( (p_1-p_2)^2 \right) f(p_1) f(p_2) d^3 p_1 d^3 p_2 \delta \left( (p_1^\mu-p_2^\mu )^2 -M^2  \right)\;\;\;;
\end{equation}
the assumption that quark and hole degrees of freedom be quark-like means $\mathcal{M}\left(...\right)$ should not be too different from the quark-quark scattering matrix elements calculated within pQCD.

The distribution functions $f(p_{1,2})$ will not however be those of a free gas but will contain a mean field reflecting the distribution of baryons in the system.   Equivalently, the Eigenstates of the unperturbed quarkyonic matter will not be free particles (i.e., the spectral function will not be flat in $M$), but rather Eigenstates of the Hamiltonian
\begin{equation}
\hat{H} = \hat{k}^2 + m_q^2 + V_{\mathrm{mf}}\left( \vec{x} \right) \eqcomma
 V_{\mathrm{mf}}\left( \vec{x},\vec{y} \right) = \sum_{i}^{\mathrm{(baryons)}} \Phi_{\mathrm{mf}}(x-x_i)
\end{equation}
where the mean field baryonic potential $\Phi_{\mathrm{mf}}$ will contain dips centered around the position of the baryons in configuration space (see Fig.~\ref{percchain}).    
If baryons were immobile and at regular intervals, as we assumed in Fig.~\ref{perc1} and \ref{perc2}, we would recover Bloch's theorem for QED conductors: the spectral function of dileptons would experience a dip at $M \simeq \rho_B^{1/3}$, as shown in Fig.~\ref{figee}, due to the disappearance of quark wavefunctions having a momentum $\simeq \rho_B^{1/3}$.   Such dips would be distinctive for quarkyonic matter, since a hadron gas would give {\em peaks} for the production of resonances and a perturbative QGP would give a relatively flat spectral function (Fig.~\ref{figee}).

Density of inhomogeneities and the movement of baryons are likely to at least somewhat spoil this picture.   However, the dynamics of baryons is calculable by semiclassical dynamical models such as uRQMD \cite{rqmd}, and shows regularities such as transverse and longitudinal expansion.   Also, experimentally we can access not just the invariant mass of the spectral function but its full vector decomposition $M_{y,r,\phi}$ in rapidity-transverse-azimuthal space.   Since in quarkyonic matter quarks are not point particles but non-trivial Eigenfunctions of the distribution of baryons, the shape of the $M_{y,r,\phi}$ distribution is likely going to be non-trivial, and reflecting macroscopic dynamics.
To calculate it, one would need to solve the Dirac equation for dips centered around a classical distribution of baryons (whose positions are given by RQMD), and then fold the solutions with pQCD interaction matrices.   This work is currently in progress, and we expect to get quantitative predictions for dilepton distributions from quarkyonic matter.

In conclusion,we have used a percolation-based model to argue for the existence of a dense ``quarkyonic'' phase in which quark wavefunctions can percolate to arbitrary distances, yet confinement persists and baryons are well-defined objects.   We have shown that this phase is almost certain to exist in the limit $N_c \gg N_f$, and is not excluded in our $N_c=3,N_f=2$ world.   We have proposed dilepton spectral functions as a possible experimental verification of this phase, and await developments at low-to-intermediate energy heavy ion collisions, such as the FAIR program.
G.~T.~acknowledges the financial support received from the Helmholtz International Center for FAIR within the framework of the
LOEWE program (Landesoffensive zur Entwicklung
Wissenschaftlich-\"Okonomischer Exzellenz) launched by the State of Hesse.

\end{document}